\documentclass[referee]{raa}
\usepackage{graphicx,times}

\begin{document}
\newcommand{\red}[1]{{\color{red} #1}}
\newcommand{\lya}{Ly$\alpha$}
\newcommand{\nv}{N\,{\footnotesize V}}
\newcommand{\siiv}{Si\,{\footnotesize IV}}
\newcommand{\civ}{C\,{\footnotesize IV}}
\newcommand{\ciii}{C\,{\footnotesize III}]}
\newcommand{\mgii}{Mg\,{\footnotesize II}}
\newcommand{\nev}{[Ne\,{\footnotesize V}]}
\newcommand{\neiii}{[Ne\,{\footnotesize III}]}
\newcommand{\hr}{H$\gamma$}
\newcommand{\hb}{H$\beta$}
\newcommand{\ha}{H$\alpha$}
\newcommand{\pr}{Pa$\gamma$}
\newcommand{\oii}{[O\,{\footnotesize II}]}
\newcommand{\oiii}{[O\,{\footnotesize III}]}
\newcommand{\hei}{He\,{\footnotesize I}}
\newcommand{\heii}{He\,{\footnotesize II}}
\newcommand{\nii}{[N\,{\footnotesize II}]}
\newcommand{\sii}{[S\,{\footnotesize II}]}
\newcommand{\feii}{Fe\,{\footnotesize II}}
\newcommand{\kmps}{$\rm km~s^{-1}$}
\newcommand{\mbh}{$M_{\rm BH}$}
\newcommand{\msun}{$M_{\odot}$}
\newcommand{\rielr}{$R_{\rm IELR}$}
\newcommand{\rsub}{$R_{\rm sub}$}

\title{Strong Optical and UV Intermediate-Width Emission Lines in the Quasar SDSS J232444.80-094600.3: Dust-Free and Intermediate-Density Gas at the Skin of Dusty Torus ?}
\author{Zhenzhen Li\inst{1,2,3,4} \and Hongyan Zhou\inst{2,1,4} \and Lei Hao\inst{3} \and Shufen Wang\inst{1,2,4} \and Tuo Ji\inst{2,4} \and Bo Liu\inst{1,2,4}}
\institute{Department of Astronomy, University of Science and Technology of China, Hefei, Anhui 230026, China; lizz08@mail.ustc.edu.cn \\
\and Polar Research Institute of China, Jinqiao Rd. 451, Shanghai, 200136, China; zhouhongyan@pric.org.cn\\
\and Key Laboratory for Research in Galaxies and Cosmology, Shanghai Astronomical Observatory, Chinese Academy of Sciences, 80 Nandan Road, Shanghai 200030, China; haol@shao.ac.cn\\
\and Key Laboratory for Research in Galaxies and Cosmology, University of Science and Technology of China, Chinese Academy of Sciences, Hefei, Anhui 230026, China}

\abstract{Emission lines from the broad emission line region (BELR) and the narrow emission line region (NELR) of active galactic nuclei (AGNs) are extensively studied. However, between these two regions emission lines are rarely detected. We present a detailed analysis of a quasar SDSS J232444.80-094600.3 (SDSS J2324$-$0946), which is remarkable for its strong intermediate-width emission lines (IELs) with FWHM $\approx$ 1800 \kmps. The IEL component is presented in different emission lines, including the permitted lines \lya\ $\lambda$1216, \civ\ $\lambda$1549, semiforbidden line \ciii\ $\lambda$1909, and forbidden lines \oiii\ $\lambda\lambda$4959, 5007. With the aid of photo-ionization models, we found that the IELs are produced by gas with a hydrogen density of $n_{\rm H} \sim 10^{6.2}-10^{6.3}~\rm cm^{-3}$, a distance to the central ionizing source of $R \sim 35-50$ pc, a covering factor of CF $\sim$ 6\%, and a dust-to-gas ratio of $\leq 4\%$ times of SMC. We suggest that the strong IELs of this quasar are produced by nearly dust-free and intermediate-density gas located at the skin of the dusty torus. Such strong IELs, served as a useful diagnose, can provide an avenue to study the properties of gas between the BELR and the NELR.
}

\authorrunning{Zhenzhen Li et al.}            
\titlerunning{Emission Gas at the Skin of the Dusty Torus}  
\maketitle

\section{INTRODUCTION}
It is generally accepted that emission lines of active galactic nuclei (AGNs) arise from two well-separated regions: the broad emission line region (BELR) and the narrow emission line region (NELR). The BELR has a smaller size (compacted within $\sim$ 1 pc) and a higher electron density ($n_{\rm e} \sim 10^9-10^{13}~\rm cm^{-3}$), generating the broad emission lines (BELs) with FWHM $\sim$ 5000 \kmps; the NELR has a larger size (extended from $\sim$ 100 pc) and a lower electron density ($n_{\rm e} \sim 10^3-10^6~\rm cm^{-3}$), giving rise to the narrow emission lines (NELs) with FWHM $\sim$ 500 \kmps.

This separation of emission-line regions yields an obvious ``gap'' between these two regions, in terms of the locations, velocity dispersion and gas densities. According to the widely accepted unified model of AGNs (e.g., Antonucci 1993), the dusty torus, located in somewhere between the BELR and the NELR, is exposed to the central ionizing source. It can be inferred that illuminated gas in the scale of the dusty torus can produce intermediate-width emission lines (IELs) through photo-ionization process. Recently, Li et al. (2015) detected prominent IELs with FWHM $\sim$ 2000 \kmps\ in a partially obscured quasar OI 287, where the conventional BELs are heavily suppressed by extinction. The clearly detected IELs provide strong evidence for the existence of the intermediate-width emission line region (IELR), which has been in debate for two decades (e.g., Wills et al. 1993; Brotherton et al. 1994a, b, 1996; Mason et al. 1996; Sulentic et al. 2000; Hu et al. 2008; Zhu et al. 2009; Zhang 2011, 2013). Detail analyses to this quasars showed that the IELs are produced by gas located in the dusty torus. Besides, emission lines originated from the dusty torus are also found in the core of Fe K$\alpha$ (Shu et al. 2010; Jiang et al. 2011; Gandhi et al. 2015; Minezake \& Matsushita, 2015a, b) and coronal lines (Rose et al. 2015a, b).

If IELs are produced by gas located in the dusty torus, they can provide an new opportunity to diagnose the gas properties of the dusty torus. For instance, the line widths of IELs may reflect the location of emission gas, assuming the gas kinematics are dominated by the gravitational force of the central black hole (Jiang et al. 2011; Li et al. 2015). Also, the line intensity ratios of IELs can be used to constrain the gas physical conditions, such as the gas density and ionization parameter. In addition, the strength of IELs may reflect the mixture of gas and dust. The IELs in normal quasars are generally suggested to be weak compared with the conventional BELs and NELs, which is explained as that dust embedded in the IELRs absorbs most of the ionizing photons, and thus suppresses the line emission (Netzer \& Laor, 1993; Mor \& Netzer, 2012). Nevertheless, gas may not always mix with dust. Finding strong IELs can be helpful for understanding the mixture between the gas and the dust.

Quasars with both rest-frame UV and optical IELs are useful to constrain the physical conditions of the IELRs. For high-redshift quasars, the rest-frame UV IELs can be obtained in a large number from the Sloan Digital Sky Survey (SDSS; York et al. 2000). Nevertheless, the rest-frame optical emission lines of these objects require to be observed in the near-infrared (NIR). The Keck II (McLean et al. 1998) 10-meter telescope spectroscopically observed the rest-frame optical emission lines of 34 quasars in the NIR, which are initially used to study the variation of the fine structure constant through \oiii\ doublets. From the SDSS-Keck sample, we found a particular quasar SDSS J232444.80-094600.3 (hereafter SDSS J2324$-$0946), which is remarkable for its strong intermediate-width components shown in different emission lines: including the permitted lines \lya\ $\lambda$1216, \civ\ $\lambda$1549, semiforbidden line \ciii\ $\lambda$1909 and forbidden lines \oiii\ $\lambda\lambda$4959, 5007. The coexistence of these different IELs provides us with an opportunity to well constrain the gas physical conditions.

This paper is organized as follows. In Section 2, we describe the observations and data reduction; in Section 3, we analyze the observational data, including the emission lines, broadband spectral energy distribution (SED); in Section 4, we discuss the physical conditions and the origin of the IELR; finally, we give a brief summary and future prospect in Section 5. Throughout this paper, we use the cosmological parameters $H_0 = 70 ~ \rm km~s^{-1} \rm Mpc^{-1}$, $\Omega_{\rm M} = 0.3$, and $\Omega_{\Lambda} = 0.7$.
\section{Observations and Data Reduction}
SDSS J2324$-$0946 was spectroscopically observed by the SDSS on October 21, 2001. The observed spectrum is shown in Panel (a) of Figure 1. It is quite similar compared with the composite quasar spectrum (Vanden Berk et al. 2001), except that SDSS J2324$-$0946 clearly shows prominent extra components in the emission lines of \lya, \civ\ and \ciii.

On September 9th, 2003, this quasar was also observed by Keck II telescope with the NIRSPEC instrument, as one target of Bahcall's proposal for studying the variation of the fine structure constant through \oiii\ $\lambda\lambda$4959, 5007 (program ID A343Ns). Four $300 s$ exposures were taken using a $42'' \times 0.38''$ slit and the NIRSPEC-5 filter. This yielded a spectral resolution of $R \sim$ 2200 and a wavelength coverage of $\lambda \sim$ 1.5--1.8 $\mu$m. Wavelength calibration was carried out using the sky light observed in the same filter. Flux calibration was performed using all B type standard stars observed on the same night with the software REDSPEC. The Keck spectrum is presented in Panel (b) of Figure 1. It is striking that \oiii\ $\lambda\lambda$4959, 5007 show broad and symmetric profiles, different from commonly observed asymmetric \oiii\ doublet with a narrow core and a blue-shifted wing (e.g., Heckman et al. 1981; Wilson \& Heckman 1985; Christopoulou et al. 1997; Tadhunter et al. 2001; V{\'e}ron-Cetty et al. 2001; Zamanov et al. 2002; Komossa \& Xu 2007; Greene \& Ho 2005; Wang et al. 2011; Zhang et al. 2011; Peng et al. 2014).

\begin{figure}[!htbp]
\centering
\includegraphics[width=0.6\textwidth, angle=0]{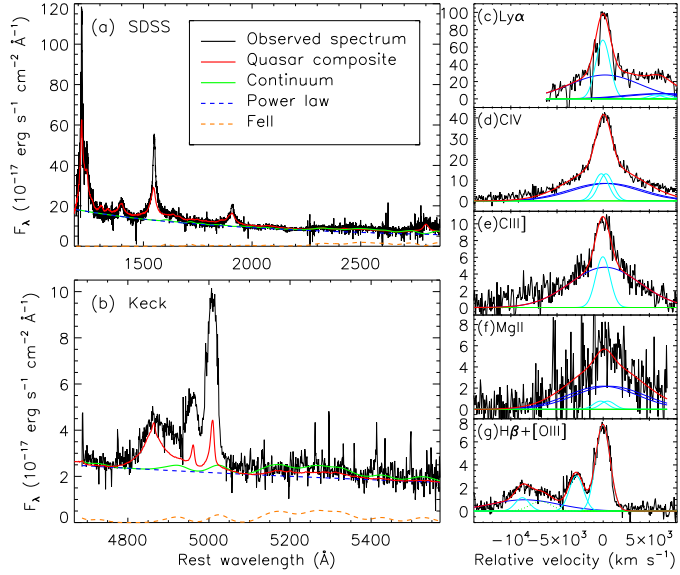}
\caption{\textbf{Left (a-b):} The observed spectra (black line) of SDSS J2324$-$0946 obtained by the SDSS and Keck. A composite quasar spectrum (red line) is overploted for a comparison. The green line denotes the continuum model, which is the sum of a single power law (blue dashed line) and a \feii\ pseudocontinuum (orange dashed line). \textbf{Right (c-g):} The emission lines (black) in their common velocity space. We decomposed the emission lines into a broad (blue) and an intermediate-width (cyan) component, assuming that the same components in different lines have the same redshift and the same profile.}
\end{figure}

\begin{table}[!htbp]
\begin{center}
\begin{minipage}[]{40mm}\caption[]{Spectroscopic Data}\end{minipage}
 \begin{tabular}{ccccccc}
  \hline
  \hline
  {Range} & {Slit}     & {$\lambda/\Delta\lambda$} & {Exp.Time} & {Instrument} & {Data} & {Reference} \\
  {(\AA)} & {(arcsec)} & {}                        & {(s)}      & {}           & {(UT)} & {}          \\
  \hline
  3800--9200    & 3.0  & 2000 & 3584 & SDSS & 2001 Oct. 21 & 1 \\
  15000--18000  & 0.38 & 2200 & 1200 & Keck & 2003 Sept. 9 & 2 \\
  \hline
\end{tabular}
\end{center}
\tablerefs{0.8\textwidth}{(1) York et al. 2003; (2) This work.}
\end{table}

\begin{figure}[!htbp]
\centering
\includegraphics[width=0.6\textwidth, angle=0]{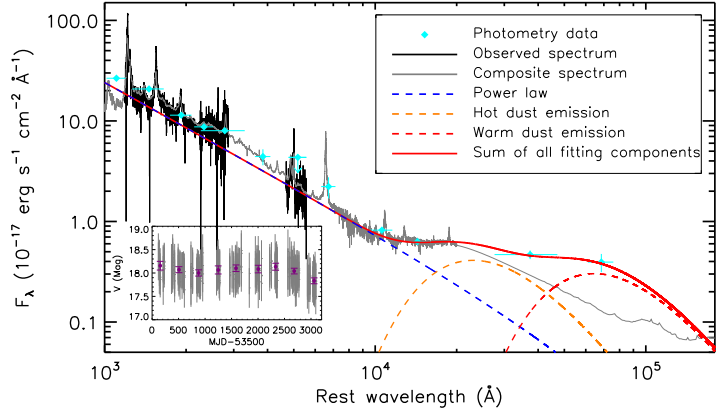}
\caption{Broadband SED of SDSS J2324$-$0946. We plot the observed photometric data (cyan diamond) and spectra (black line). The composite quasar spectrum (gray line) normalized at WISE-$W$2 is overplotted. The SED is modelled using a power law (blue dashed line), a hot (orange dashed line) and a warm (red dashed line) black body. The inserted panel shows the light curve of SDSS J2324$-$0946 at $V$ band monitored by the Catalina Sky Survey. The red dots represent the 1-$\sigma$ dispersion for each season.}
\end{figure}

\begin{table}[!htbp]
\begin{center}
\begin{minipage}[]{0.43\textwidth}\caption[]{~~~~~~~~~~~~Photometric Data}\end{minipage}
 \begin{tabular}{lcclc}
  \hline
  \hline
  {Band}  & {Value}  & {Facility} & {Date} & {Reference} \\
  {}      & {(mag)}  & {}         & {(UT)} & {}          \\
  \hline
$u$  & 18.959$\pm$0.029 & SDSS  & 2000 Nov. 17 & 1 \\
$g$  & 18.552$\pm$0.021 & SDSS  & 2000 Nov. 17 & 1 \\
$r$  & 18.575$\pm$0.022 & SDSS  & 2000 Nov. 17 & 1 \\
$i$  & 18.426$\pm$0.023 & SDSS  & 2000 Nov. 17 & 1 \\
$z$  & 18.102$\pm$0.034 & SDSS  & 2000 Nov. 17 & 1 \\
$J$  & 17.150$\pm$0.181 & 2MASS & 1998 Oct. 11 & 2 \\
$H$  & $<$16.036        & 2MASS & 1998 Oct. 11 & 2 \\
$K_s$& 15.724$\pm$0.231 & 2MASS & 1998 Oct. 11 & 2 \\
$W1$ & 14.984$\pm$0.035 & WISE  & 2010 Jun. 8  & 3 \\
$W2$ & 13.970$\pm$0.044 & WISE  & 2010 Jun. 8  & 3 \\
$W3$ & 10.374$\pm$0.102 & WISE  & 2010 Jun. 8  & 3 \\
$W4$ &  7.798$\pm$0.203 & WISE  & 2010 Jun. 8  & 3 \\
  \hline
\end{tabular}
\end{center}
\tablerefs{0.6\textwidth}{(1) York et al. 2003; (2) This work.}
\end{table}

We also collected photometric data of this quasar from the SDSS, Two Micron All Sky Survey (2MASS; Skrutskie et al. 2006) and Wide-Field Infrared Survey Explorer (WISE; Wright et al. 2010). Figure 2 shows the constructed broadband spectral energy distribution (SED). In addition, we investigated the light curve of the \textit{V}-band magnitude monitored by the Catalina Sky Survey\footnote{http://nesssi.cacr.caltech.edu/DataRelease/} for nearly 7 years (2006/09/22--2013/07/27, insert panel of Figure 2). The variation amplitudes of this quasar are within 0.15 magnitude, indicating that the variation is insignificant.

Before further analysis, all of the spectroscopic and photometric data have been corrected for a Galactic reddening of $E(B-V)$=0.028 using the updated dust map of Schlafly \& Finkbeiner (2011) and converted to the rest frame of the quasar using the redshift $z=2.2116$ \footnote{The redshift is determined by the peaks of low-ionized lines \hb\ and \mgii.}.
\section{Data Analysis and Results}
\subsection{Emission Lines}
The most significant feature in the observed spectrum of SDSS J2324$-$0946 is the intermediate-width and symmetric \oiii\ lines $\lambda\lambda$4959, 5007. Generally, the \oiii\ doublet in the spectra of AGNs are narrow lines. In particular case, \oiii\ lines can be broadened by outflows. This usually gives a narrow core and blue-shifted wing. However, the \oiii\ lines in SDSS J2324$-$0946 is rather symmetric. This profiles is also shown in the top of permitted BELs, especially prominent in \lya, \civ, and \ciii. In order to model these line profiles better, we first take out the continuum of the SDSS and Keck spectrum using a single power law and a \feii\ pseudocontinuum in the continuum windows. After subtracting the continuum model from the observed spectrum, we obtain strong emission lines of \lya, \civ, \ciii, \mgii, \hb\ and \oiii\ shown in the right panels of Figure 1.

The profiles of the \oiii\ doublet are quite symmetric. To test this quantitatively, we fit each of the \oiii\ lines with two models: one uses a single Gaussian and the other uses two Gaussains. By comparing the two fitting models with a F-test, we found that the \oiii\ lines cannot be improved significantly using the two Gaussain model with a chance probability less than 0.05. This indicates that a singe Gaussian is enough for \oiii\ lines.

The other emission lines, including \lya, \civ, \ciii, \mgii\ and \hb, also show a similar intermediate-width component as \oiii. Besides, these emission lines contain a broad component. We also separately fit these emission lines using a single Gaussian in one model and two Gaussians in another model, and compare these two models with a F-test. The result shows that these emission lines need two Gaussians. We decompose these lines into a broad and an intermediate-width components. Each component is modeled with a single Gaussian. The broad components in different lines are assumed to have the same redshift and line width. The intermediate-width components are forced to have the same redshif and line width as \oiii\ lines.

For the doublets of \civ\ and \mgii, each doublet component is fitted separately with their relative intensity ratios fixed at 1:1. The \hb\ shows an excess in the red wing (the ``red shelf'', Meyers \& Peterson 1985; V{\'e}ron et al. 2002). Detailed study of this feature is beyond the scope of this paper, and we use an additional Gaussian to eliminate its influence. Absorption lines and bad pixels are carefully masked. We simultaneously fit all of these emission lines using an Interactive Data Language (IDL) code based on MPFIT (Markwardt 2009), which performs $\chi^2$--minimization by the Levenberg Marquardt technique. The best-fit results are shown in the right panels of Figure 1. The emission lines can be well modelled with a broad component with FWHM of $8940\pm167$ \kmps\ and an intermediate-width component with FWHM of $1832\pm26$ \kmps. The emission-line parameters are summarized in Table 3.
\begin{table}[!htbp]
\begin{center}
\begin{minipage}[]{0.4\textwidth}\caption[]{Measurements of Emission Lines}\end{minipage}
 \begin{tabular}{ccc ccccccc}
  \hline
  \hline
  {Component}&{Shift}  &{FWHM}   &\multicolumn{7}{c}{Flux} \\
  {}         &{(\kmps)}&{(\kmps)}&\multicolumn{7}{c}{($\rm 10^{-17}~erg~s^{-1}~cm^{-2}$)} \\
  \cline{4-10}
  {}         &{}       &{}       &{\lya}&{\civ}&{\ciii}&{\mgii}&{\hb}&{\oiii$\lambda$4959}&{\oiii$\lambda$5007}\\
  \hline
  BELs&     199$\pm      45$&    8940$\pm     167$&    3421$\pm     174$&    2660$\pm      65$&     935$\pm      61$&    1242$\pm     107$&     483$\pm      56$&--&--\\
  IELs&     -36$\pm      10$&    1832$\pm      26$&    1725$\pm      92$&     845$\pm      48$&     241$\pm      28$&      89$\pm      54$&      94$\pm      26$&     224$\pm      20$&     623$\pm      51$\\
  \hline
\end{tabular}
\end{center}
\end{table}

\subsection{Broadband SED}
As shown in Figure 2, the broadband SED of SDSS J2324$-$0946 is identical to the quasar composite spectra (Vanden Berk et al. 2001) from the rest-frame ultraviolet (UV) to NIR, but has an obvious excess in the mid-infrared (MIR). It is commonly believed that quasar SEDs in the NIR and MIR are dominated by the thermal radiation of hot and warm dust (e.g., Polletta et al. 2000; Klaas et al. 2001; Nenkova et al. 2002). The MIR excess of this quasar implies a larger warm dust radiation. We decompose the SED of SDSS J2324$-$0946 into three components: a single power law to mimic the emission from the accretion disk, and two black body for the thermal radiation of the hot and warm dust. The results are shown in Figure 2.

The ratio of the infrared luminosity to bolometric luminosity, $L_{\rm IR}/L_{\rm bol}$, is generally interpreted as an estimator of the dust covering factor (CF) (Maiolino et al. 2007; Hatziminaoglou et al. 2008; Rowan-Robinson et al. 2009; Roseboom et al. 2013). From the observed SED of SDSS J2324$-$0946, we derive the NIR and MIR luminosity, $L_{\rm NIR} = 1.7\pm0.3 \times 10^{46}~\rm erg~s^{-1}$ and $L_{\rm MIR} = 3.5\pm0.5 \times 10^{46}~\rm erg~s^{-1}$. Using the continuum luminosity at 5100 \AA\ ($L_{\lambda}(5100)$ \AA) and the bolometric correction $L_{\rm bol}=9 \lambda L_{\lambda}(5100)$ \AA\ (Kaspi et al. 2000), the bolometric luminosity of this quasar is evaluated to be $L_{\rm bol} \sim 1.1 \times 10^{47}~\rm erg~s^{-1}$. With these estimations, the hot and warm dust CFs are derived to be ${\rm CF}_{\rm HD}=15\%$, and ${\rm CF}_{\rm WD}=31\%$, respectively. Compared with the measurement results (${\rm CF}_{\rm HD}=15\%$, ${\rm CF}_{\rm WD}=23\%$) for a large sample of type I quasars (Roseboom et al. 2013), SDSS J2323-046 has a typical hot dust CF and a larger warm dust CF.

With the estimated dust CF, the dust luminosity can be expressed as $L_{\rm dust}=4 \pi R_{\rm dust}^2 {\rm CF} \sigma T^4$, where $\sigma$ is the Stefan-Boltzmann constant, $R_{\rm dust}$ is the distance of the dust to the central source, and $T$ is the dust temperature. From the SED decomposition, the hot and warm dust temperatures are estimated to be $T_{\rm HD} \sim 1260$ K and $T_{\rm WD} \sim 450$ K. These yield estimations of $R_{\rm HD} \sim 2.5$ pc and $R_{\rm WD} \sim 20$ pc. These estimations will be used to study the origin of the IELs in Section 4.2.
\section{DISCUSSION}
\subsection{Physical Conditions of the Intermediate-width Emission Line Region}
With the measurements of the line intensities of various emission lines, we investigate the physical conditions of the IELR, using photo-ionization model calculations. We consider a gas slab with solar abundance, which is illuminated by an ionizing source with an SED defined by Mathews \& Ferlan (1987, hereafter MF87). For simplicity, we assume that the gas is dust free and ionization bound\footnote{This is done by setting a large enough total hydrogen column density of $N_{\rm H}=10^{24}~\rm cm^{-2}$.}. The justification of these assumptions will be discussed later in detail. We perform the models using the CLOUDY code (Version 13.03, Ferland et al. 1998).

We calculate a two-dimensional grid with variable hydrogen density ($n_{\rm H}$) in the range of $10^{3}-10^{9}~\rm cm^{-3}$ and ionization parameter ($U$) in the range of $10^{-3}-10^{0}$. Both of $n_{\rm H}$ and $U$ vary with a small step of 0.1 dex. The calculated results are shown in Figure 3, where we plot the contours of line-intensity ratios (compared to the \civ\ flux). We use the observed IELs in SDSS J2324$-$0946 to constrain the parameters. The colored areas represent the observed ranges for 1-$\sigma$ measurement errors. Most of the line-intensity ratios intersect at area of $n_{\rm H} \sim 10^{6.2}-10^{6.3}~\rm cm^{-3}$ and $U \sim 10^{-1.6}-10^{-1.4}$.

\begin{figure}[!htbp]
\centering
\includegraphics[width=0.8\textwidth, angle=0]{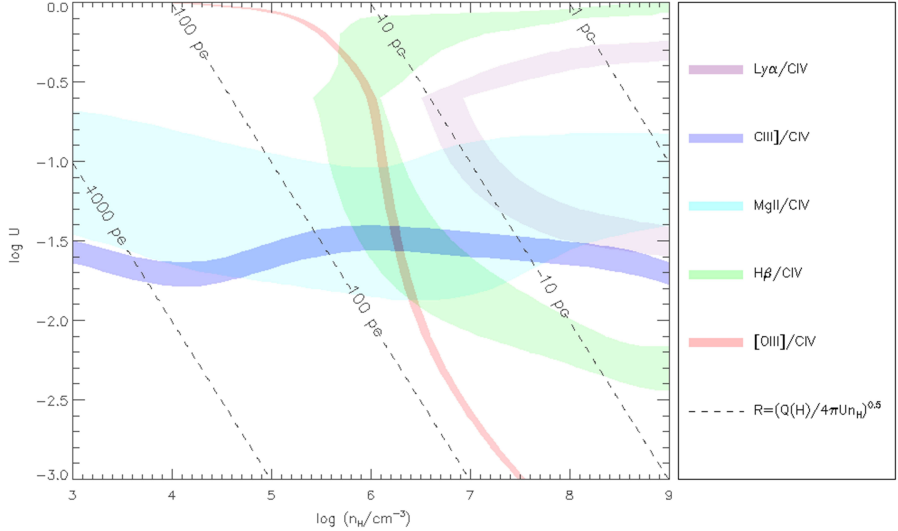}
\caption{Contours of line-intensity ratios (normalized by \civ) as functions of $n_{\rm H}$ and $U$. The model is calculated in the case: MF87 SED, solar abundance, dust free, and ionization bound. The filled areas represent the observed ranges for 1-$\sigma$ measurement errors. Most of the line-intensity ratios form an overlap of $n_{\rm H} \sim 10^{6.2}-10^{6.3}~\rm cm^{-3}$ and $U \sim 10^{-1.6}-10^{-1.4}$. Within these parameter ranges, the emission-line region to the central ionizing source is constrained in the range of $R \sim 35-50$ pc, as indicated by the gray dashed lines, which represent $R=(Q({\rm H})/4\pi c U n_{\rm H})^{0.5}$.}
\end{figure}

With the estimated $n_{\rm H}$ and $U$ above, the distance of the emitting region to the central ionizing source can be derived as $R=(Q({\rm H})/4\pi c U n_{\rm H})^{0.5}$, where $Q(\rm H)$ is the number of hydrogen ionization photons, $Q({\rm H})=\int_{\nu}^{\infty} L_{\nu}/h \nu d \nu \approx 3.5 \times 10^{56}~\rm photons~s^{-1}$. In Figure 3, we also show the contours (dotted lines) of $R$ as functions of $n_{\rm H}$ and $U$. In the overlapping region, $R$ is constrained to be in the range of $R \sim 35-50$ pc. By combining $R$ with the IEL width and assuming that the IELR is virialized, the black hole mass (\mbh) of this quasar can be estimated as, $M_{\rm BH}=R (f {\rm FWHM(IEL)})^2/G$, where G is the gravitational constant and $f$ is a scaling factor. With a simple approximation of an isotropic IELR and Gaussian-profile IELs, $f=\sqrt{3}/2.354$\footnote{For an isotropic IELR, the velocity dispersion ($\sigma)$ along the line-of-sight ($\sigma_{\rm line}$) is equal in all directions, $\sigma=\sqrt{3} \sigma_{\rm line}$. For a Gaussian profile of IELs, $\sigma_{\rm line}=\rm {FWHM(IELs)}/2.354$. Thus, the scale factor is $f \equiv \sigma/\rm{FWHM(IELs)} = \sqrt{3}/2.354$.} (Li et al. 2015). With these assumptions, we derive an estimate of \mbh\ $\sim 1.5-2.1 \times 10^{10}$ \msun. This is roughly consistent with the value of $M_{\rm BH} \sim 8.7 \times 10^8 - 2.5 \times 10^{10}$ \msun, estimated using the \mgii\ broad line width and continuum luminosity at 3000 \AA, and employing the empirical formula in Wang et al. (2009). The agreement indicates that the kinematics of IELR clouds, the same as that of the BELR, is dominated by the gravity force of the central black hole.
\begin{figure}[!htbp]
\centering
\includegraphics[width=0.6\textwidth, angle=0]{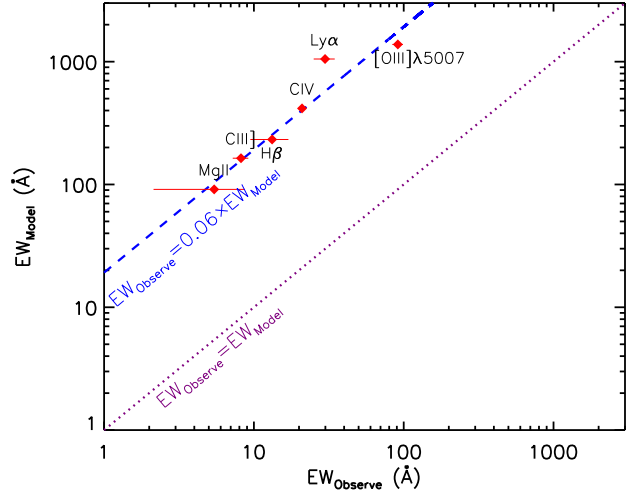}
\caption{Comparison between model-predicted EWs and observed EWs. The observed EW are smaller than their model values for an approximate scale. The purple doted line represents a 1:1 relationship of ${\rm EW}_{\rm observe}={\rm EW}_{\rm model}$. The blue dashed line represents the best-fit relation of ${\rm EW}_{\rm observe} = 0.06 \times {\rm EW}_{\rm model}$.}
\end{figure}

Adopting the parameters constrained above, we predict the equivalent widths (EWs) of all strong IELs in SDSS J2324$-$0946. Figure 4 shows the comparison between the mode-predicted EWs (${\rm EW}_{\rm model}$) and the observed EWs (${\rm EW}_{\rm observe}$). A linear fit yields a relationship of ${\rm EW}_{\rm observe} = 0.06 \times {\rm EW}_{\rm model}$ (blue dashed line). The purple dotted line denotes the relationship of ${\rm EW}_{\rm observe}={\rm EW}_{\rm model}$. It is clearly seen that all of the observed EWs are smaller than model EWs for an approximate value. The smaller observed EWs can be naturally explained by the gas covering factor, as the model is calculated in the case of full coverage, and all emission line EWs are proportional to the value of gas covering factor. Thus, the ratio of ${\rm EW}_{\rm model}$ to ${\rm EW}_{\rm observe}$ can be interpreted as an estimator of the gas covering factor, as ${\rm CF}={\rm EW}_{\rm observe}/{\rm EW}_{\rm model}=6\%$.

The observed \lya/\civ\ in Figure 3 and observed EW(\lya) in Figure 4 are in a slightly smaller location. This may be explained by serval factors. (1) The \lya\ emission line might be absorbed by neutral hydrogen, as indicated by the absorption lines shown in the observed SDSS spectrum (Figure 1), while there is no clear absorption lines in the other lines. (2) Compared with other lines, the resonant \lya\ photons in the line-of-sight are easier to be scattered by neutral hydrogen into other directions. This could lower the intensity of \lya. (3) Another effect due to the resonant property of \lya\ is that \lya\ is easier to be attenuated by dust extinction than the other lines. We carry out an additional photo-ionization simulations by adding Small Magellanic Cloud (SMC)-like grain as an example to investigate the effect of dust. The dust-to-gas ratio increases from 0.1 to 0.4 times of SMC. For each dust-to-gas ratio, we repeat the dust-free process as described above. With the modeled and observed five line intensity ratios, we derive a best-fit $\chi^2=\sum_{i=1}^5((observe_i-model_i)/error_i)^2$, where $observe_i$, $model_i$ and $error_i$ is one of the observed, modeled and measurement error of the line intensity ratios. The left panel of Figure 5 shows the relation between the best-fit reduced $\chi^2$ ($\chi_r^2$) and the dust-to-gas ratio. The $\chi_{r}^2$ is minimal when dust-to-gas ratio is only 4\% SMC. The dust-to-gas ratio of SMC is low, about $1/20$ of the Milky Way. Thus, the emission gas of SDSS J2324$-$0946 is very dust-poor, even if the smaller \lya\ is caused by dust.

The above simulations are calculated in the case of ionization bound. To demonstrate the rationality of this assumption, we also carry out a photo-ionization model by varying the $N_{\rm H}$ from a small value of $10^{19}~\rm cm^{-2}$ to a vary large value of $10^{24}~\rm cm^{-2}$. For each $N_{\rm H}$, we repeat the ionization bound process as described above and derive a best-fit $\chi_{r}^2$. The right panel of Figure 5 shows the relation between the best-fit $\chi_{r}^2$ with $N_{\rm H}$. It is clearly shown that $\chi_{r}^2$ decrease quickly with increasing $N_{\rm H}$ in the beginning. When $N_{\rm H}$ is larger than $10^{22.5}~\rm cm^{-2}$, $\chi_{r}^2$ approach to a constant value. This indicate that the ionizing bound condition in SDSS J2324$-$0946 can well produce the observed lines.

\begin{figure}[!htbp]
\centering
\includegraphics[width=0.8\textwidth, angle=0]{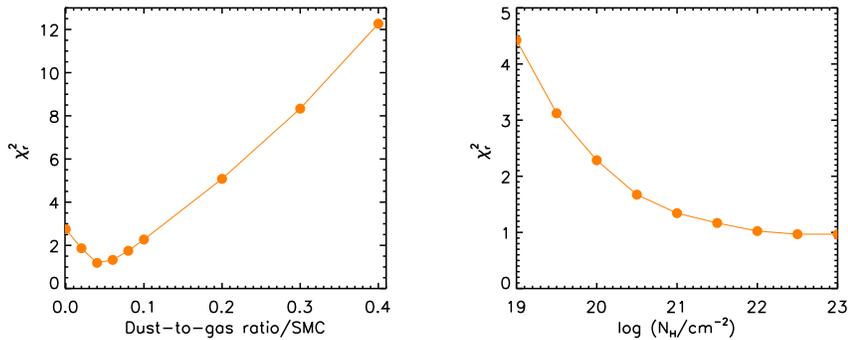}
\caption{\textbf{Left:} The relation between the best-fit $\chi_r^2$ and the dust-to-gas ratio. The $\chi_r^2$ is minimal when the dust-to-gas ratio is only 4\% times of SMC. \textbf{Right:} The relationship between the best-fit $\chi_r^2$ and column density. The $\chi_r^2$ decrease quickly at the beginning, and approach to a constant value when $N_{\rm H}>10^{22.5}~\rm cm^{-2}$.}
\end{figure}

\subsection{The origin of the Intermediate-Width Emission Lines}
The IELs of SDSS J2324$-$0946 may not originate from the typical NELR. It is generally believed that NELs can serve as a surrogate for the stellar velocity dispersion ($\sigma_{*}$). However, the line width of IELs in SDSS J2324$-$0946 is $\sigma \approx$ 800 \kmps, which is much larger than the largest value of $\sigma_{*} = 444$ \kmps\ for known galaxies (Salviander et al. 2008). Also, according to the $M_{\rm BH}-\sigma_{*}$ relationship of Tremaine et al. (2002), $\sigma_{*}$ is derived to be $\sim$ 420 \kmps, obviously smaller than those of the IELs.

It is also unlikely that the IELs originate from the BELR. On one hand, the BELRs generally have a large density ($\sim 10^{10}~\rm cm^{-3}$, Netzer et al. 2013). To estimate the physical conditions of BELR in this object, we repeat the CLOUDY calculations following the processes of the IELR described above. As shown in Figure 6, most of the BEL intensity ratios form an overlap of $n_{\rm H} \sim 10^{10.1}-10^{10.3}~\rm cm^{-3}$ and $U \sim 10^{-2.1}-10^{-2.2}$. The \lya/\civ\ of BELs is away from the overlap (the same as that of IELs), which may also caused by those factors mentioned in Section 4.1. Since the estimated density for the BELR in this quasar is much larger than the critical density of \oiii$\lambda$5007, $n_{\rm crit} = 7 \times 10^5~\rm cm^{-3}$, the observed strong \oiii\ $\lambda$5007 is hard to be produced in the BELR. On the other hand, the BELR volume is too small to produce the strong IELs. Assuming that the BELR of SDSS J2324$-$0946 is fully filled by gas with a density of the critical density of \oiii\ $\lambda$5007, the predicted \oiii\ $\lambda$5007 luminosity is estimated to be $L_{\rm \oiii \lambda5007} = 4/3 \pi R_{\rm BELR}^3 j_{\rm \oiii \lambda5007}$, where the BELR radius is estimated to be $R_{\rm BELR} \sim$ 0.5 pc using the radius-luminosity relation of BELR (Kaspi et al. 2005), and the \oiii $\lambda$5007 volume emissivity is $j_{\rm \oiii \lambda5007} \sim 10^{-13.6}~\rm erg~s^{-1}~cm^{-3}~ster^{-1}$ from CLOUDY calculations with the critical density of \oiii\ $\lambda$5007. These estimations yield $L_{\rm \oiii\ \lambda5007} \sim 4.3 \times 10^{41}~\rm erg~s^{-1}$, which is much less than the observed value of $2.2 \times 10^{44}~\rm erg~s^{-1}$.

\begin{figure}[!htbp]
\centering
\includegraphics[width=0.8\textwidth, angle=0]{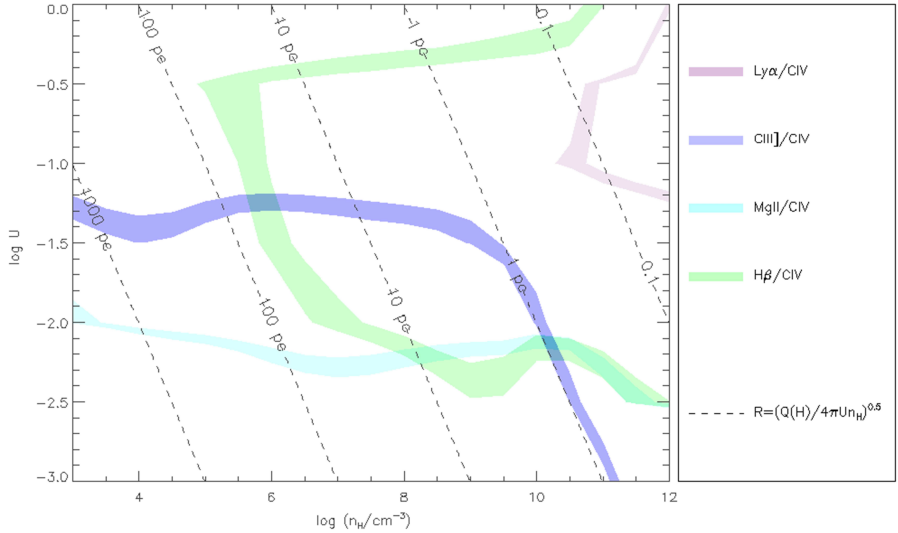}
\caption{Same as Figure 3, but for the BELR. Most of the line-intensity ratios form an overlap of $n_{\rm H} \sim 10^{10.1}-10^{10.3}~\rm cm^{-3}$ and $U \sim 10^{-2.1}-10^{-2.2}$.}
\end{figure}

Intermediate-width \oiii\ emission lines are also observed in some radio-loud quasars, which is believed to be generated by jet-induced outflow (Kim et al. 2013). This is also unlikely for SDSS J2324$-$0946 since its radio intensity is very faint, not detected by the NRAO VLA Sky Survey (NVSS; Condon et al. 1998) with a detection limit of 2.5 mJy at 1.4 GHZ.

Based on the photo-ionization model calculations in Section 4.1, we found that the IELR of this quasar has a hydrogen density of $n_{\rm H} \sim 10^{6.2}-10^{6.3}~\rm cm^{-3}$ and a distance to the central ionizing source of $R \sim 35-50$ pc. Both of the inferred $n_{\rm H}$ and $R$ suggest that the IELR of this quasar may originate from somewhere of the dusty torus. The inside part of the dusty torus is unlikely to produce the observed IELs. In this region, clouds are hard to be illuminated and emission lines are difficult to escape from the dusty torus, even if the dusty torus is clumpy (e.g., Krolik \& Begelman 1988). The places far away from the dusty torus (e.g., the pole-on regions) are also unlikely to produce the observed IELs in this quasar, as clouds in these regions are easy to be accelerated by the radiation pressure. Many researches reported that a large fraction ($\sim$ 50\%) of Seyfert 1 galaxies have blue-shifted intrinsic UV absorption lines (e.g., Anderson \& Kraft, 1969; Crenshaw et al. 1999; Kriss 2002; Dunn et al. 2007, 2008; Ganguly \& Brotherton 2008), which suggests that outflow should be common in the ionization cone. It can be inferred that emission lines produced in this region should have prominent blue shifts, which is inconsistent with the observed IELs in SDSS J2324$-$0946. The skin of the dusty torus may be the most reasonable location producing the IELs in this object. In this region, clouds can be illuminated by the central ionization source and emission lines can easily escape. Besides, the dusty torus could be a reservoir for supplying clouds onto the skin. There can be a great deal of clouds for producing emission lines at the skin of the dusty torus. As the kinematics of these clouds should be similar to those of the dusty tours, the redshifts of the produced emission lines can be consistent with the systemic redshift of the quasar.

The photo-ionization model inferred $n_{\rm H}$ of this quasar is much lower than the typical gas density near the inner part of the dusty torus, $\sim 10^9~\rm cm^{-3}$ suggested by recent observations (Kishimoto et al. 2013) and modeling (Stern et al. 2014). In addition, according to the radius-luminosity relation of dusty torus (Koshida et al. 2014), the inner radius of the dusty torus in SDSS J2324$-$0946 is estimated to be ~2 pc. This value is obviously smaller than the radius of the IEL emitting region, $R \sim 35-50$ pc. The comparison indicates that the IEL emission region of this quasar may locate in a far part of the dusty torus. In this location, illuminated gas with an intermediate density can produce both permitted and forbidden emission lines, as observed in the spectrum of SDSS J2324$-$0946.

The IELs in typical quasars are generally suggested to be very weak, since dust mixed in the IELRs can suppress the line emission (Netzer \& Laor, 1993; Mor \& Netzer, 2012). The photo-ionization model shows that dust in the IELR of SDSS J2324$-$0946 is very poor. This may be the reason of the strongness of IELs shown in this quasar. The strong IELs of SDSS J2324$-$0946 imply that the mixture of dust and gas may be not uniformity in the dusty torus. There maybe also gas, which is not mixed with dust, locate in the dusty torus. This gas, illumined by the central ionizing source, can produce strong IELs through photo-ionization process. Quasars with strong IELs, such as SDSS J2324$-$0946, can help us to investigate the physical condition of this gas.

\section{Summary and Future Prospect}
With the SDSS and Keck observations of the quasar SDSS J2324$-$0946, we presented a detailed analysis of its emission lines. The emission lines is remarkable for its strong IELs with FWHM $\approx$ 1800 \kmps\ shown in various lines, including the permitted lines \lya\ $\lambda$1216, \civ\ $\lambda$1549, semiforbidden line \ciii\ $\lambda$1909, and forbidden lines \oiii\ $\lambda\lambda$4959, 5007. The coexistence of these different IELs provides us with an opportunity to constrain the gas physical conditions. With the measurements of the IELs, we investigate the physical conditions of emission gas, using photo-ionization model calculations. We found that the IELs are produced by gas with a hydrogen density of $n_{\rm H} \sim 10^{6.2}-10^{6.3}~\rm cm^{-3}$, an ionization parameter of $U \sim 10^{-1.6}-10^{-1.4}$, a distance to the central ionizing source of $R \sim 35-50$ pc, a covering factor of CF $\sim$ 6\%, and a dust-to-gas ratio of only $4\%$ times of SMC at most. We discussed the origin of the IELR, and found that the IELs of this quasar are unlikely from the NELR, BELR, nor jet-induced outflow. We suggest that the strong IELs of this quasar are produced by nearly dust-free and intermediate-density gas located at the skin of the dusty torus. The case study of SDSS J2324$-$0946 suggests that there are also gas emission lines from a location between the conventional BELR and NELR of AGNs. Quasars with strong IELs, such as SDSS J2324$-$0946, can help us to evaluate the physical properties of these gas.

The IELR of SDSS J2324$-$0946 is suggested to has a density of $n_{\rm H} \sim 10^{6.2}-10^{6.3}~\rm cm^{-3}$. If this is correct, it is expected that the IELs of this quasar would be appear in more emission lines, especial those forbidden lines with a larger critical density, such as \nev\ $\lambda$3426 ($n_{\rm crit}=10^{7.20}~\rm cm^{-3}$), \neiii\ $\lambda$3869 ($n_{\rm crit}=10^{6.99}~\rm cm^{-3}$), and \oiii\ $\lambda$4363 ($n_{\rm crit}=10^{7.52}~\rm cm^{-3}$). In the meantime, forbidden lines with smaller critical density, such as \oii\ $\lambda$3727 ($n_{\rm crit}=10^{3.65}~\rm cm^{-3}$), \sii\ $\lambda$6583 ($n_{\rm crit}=10^{3.18}~\rm cm^{-3}$), and \sii\ $\lambda$6731 ($n_{\rm crit}=10^{3.59}~\rm cm^{-3}$) would not show IELs. These emission lines, not covered by SDSS and Keck spectrum in this work, deserve to be checked by further spectroscopically observations.

SDSS J2324$-$0946 is not unique among SDSS quasars. This quasar has a \mbh\ of $\sim 4.6 \times 10^9$ \msun, a bolometric luminosity of $\sim 1.1 \times 10^{47}~\rm erg~s^{-1}$, and an Eddington ratio of $\sim$ 0.3, all of which are normal in the sample of SDSS DR7 quasars (Shen et al. 2011). Besides, the UV/optical spectral index of this quasar is also typical, as its SED is nearly identical to the quasar composite spectrum from the rest-frame UV to NIR (Figure 2). These comparisons suggest that the appearance of IELs in SDSS J2324$-$0946 are not relate to the properties of the central engine, but to the local environment of the emission region. Finding more similar objects could help us to understand the generation conditions of strong IELs.

\begin{figure}[!htbp]
\centering
\includegraphics[width=0.8\textwidth, angle=0]{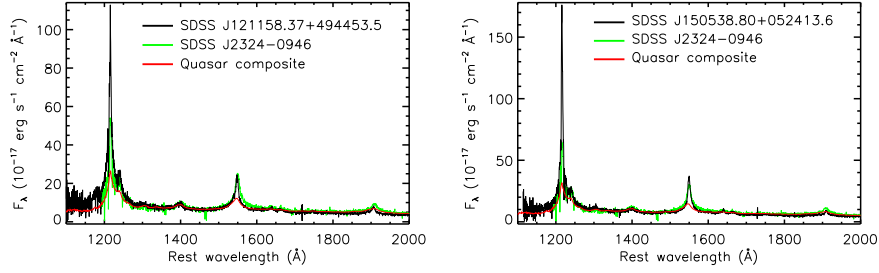}
\caption{Examples of candidates with strong UV intermediate-width component found from BOSS quasar catalog (black). The spectrum of SDSS J2324$-$0946 (green) and the composite quasar spectrum (red; Vanden Berk et al. 2001) is overploted for comparison.}
\end{figure}

\begin{figure}[!htbp]
\centering
\includegraphics[width=0.8\textwidth, angle=0]{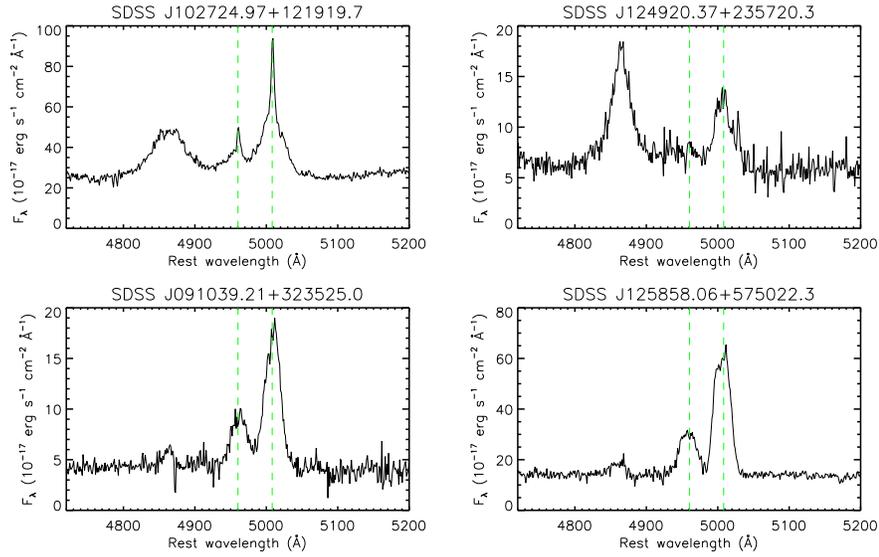}
\caption{Examples of candidates with intermediate-width \oiii\ found from type I (top panels) quasar catalog (Schneider et al. 2010) and type II (bottoms panels) quasar catalog (Reyes et al. 2008).}
\end{figure}

As mentioned earlier, SDSS J2324$-$0946 was found from a sample of 34 quasars with both rest-frame UV and optical emission line spectra. From this small sample, we did not find any other objects similar to SDSS J2324-0946. Taking SDSS J232-0946 as a prototype, we find analogues of SDSS J2324$-$0946 from large spectral sky surveys. Firstly, we search SDSS J2324$-$0946 analogues in high redshift through finding objects with similar rest-frame UV IELs. From BOSS DR12 quasar catalog (Alam et al. 2015) in the redshifts of $2<z<2.5$ ($\sim 10^5$ objects), we preliminarily found tens of thousands objects whose UV emission lines present a strong intermediate-with component. The large number of these objects shows that UV IELs can be widely detected in the spectra of quasar, as noted by previous studies (e.g., Wills 1993; Brotherton 1994). Figure 7 shows examples of selected candidates. Besides, we also search SDSS J2324$-$0946 analogues in the low redshift through finding objects with similar intermediate-width \oiii\ doublet. From SDSS DR7 quasar catalog (Schneider et al. 2010) in the redshifts of $z<0.8$ ($\sim 20,000$ objects), we found about 150 objects whose with intermediate-with \oiii. The top panels of Figure 8 shows examples of selected objects. It is clearly seen that these \oiii\ lines contain an obvious intermediate-width component. In addition, the analyzed results of SDSS J2324$-$0946 show that \oiii\ IELs are produced in a large region with a distance to the central black hole for dozens of parsec. According to the unified model of AGNs (e.g., Antonucci 1993), gas in this region can also be observed for type II quasar. Since the central BELR is obscured in these quasars, the IELs can avoid the uncertainties of line decomposition from broad \hb\ and \feii. From a 887 type II quasar catalog (Reyes et al. 2008), we also found 6 type II quasars with strong intermediate-width \oiii. The fraction (0.5\%) is consistent with that of type I sample (0.6\%). The bottom panels of Figure 8 show examples of selected objects. These analogues found above demonstrate that SDSS J2324$-$0946 is not unique. The details of the sample selection will be described in a forthcoming paper. Further studies of these objects could help us to understand gas between the conventional BELR and NELR.

We thank the anonymous referee for careful comments and helpful suggestions that led to the improvement of the paper. This work is supported by the SOC program (CHINARE 2012-02-03), Natural Science Foundation of China grants (NSFC 11473025, 11033007, 11421303, 11503022, 11473305), National Basic Research Program of China (the 973 Program 2013CB834905), and Strategic Priority Research Program ``The Emergence of Cosmological Structures'' (XDB 09030200).

Funding for SDSS and SDSS-II has been provided by the Alfred P. Sloan Foundation, Participating Institutions, National Science Foundation, U.S. Department of Energy, NASA, Japanese Monbukagakusho, Max Planck Society, and Higher Education Funding Council for England. The SDSS is http://www.sdss.org/.

SDSS is managed by the Astrophysical Research Consortium for the Participating Institutions. The Participating Institutions are the American Museum of Natural History, Astrophysical Institute Potsdam, University of Basel, University of Cambridge, Case Western Reserve University, University of Chicago, Drexel University, Fermilab, Institute for Advanced Study, Japan Participation Group, Johns Hopkins University, Joint Institute for Nuclear Astrophysics, Kavli Institute for Particle Astrophysics and Cosmology, Korean Scientist Group, Chinese Academy of Sciences (LAMOST), Los Alamos National Laboratory, Max-Planck-Institute for Astronomy (MPIA), Max-Planck-Institute for Astrophysics (MPA), New Mexico State University, Ohio State University, University of Pittsburgh, University of Portsmouth, Princeton University, United States Naval Observatory, and the University of Washington.


\end{document}